\documentclass[aps,prl,reprint,twocolumn,superscriptaddress,longbibliography,nofootinbib,floatfix,showpacs]{revtex4-1}

\usepackage{graphicx,color}
\usepackage{amsmath,amssymb}
\usepackage{hyperref}

\long\def\symbolfootnote[#1]#2{\begingroup%
\def\thefootnote{\fnsymbol{footnote}}\footnote[#1]{#2}\endgroup} 
\def\blfootnote{\xdef\@thefnmark{}\@footnotetext}

\definecolor{purp}{rgb}{0.5,0,0.5}
\definecolor{darkgreen}{rgb}{0.1,0.7,0}
\definecolor{orange}{rgb}{1,0.6,0}

\newcommand{\be}{\begin{equation}}
\newcommand{\ee}{\end{equation}}
\newcommand{\bi}{\begin{itemize}}
\newcommand{\ei}{\end{itemize}}
\newcommand{\bea}{\begin{eqnarray}}
\newcommand{\eea}{\end{eqnarray}}

\newcommand{\E}{\mathbf{E}}
\newcommand{\BB}{\mathbf{B}}

\begin{document}

\title{Anomalous radiative trapping in laser fields of extreme intensity}

\author{A.~Gonoskov}
\email[]{arkady.gonoskov@physics.umu.se}
\affiliation{Department of Physics, Ume\aa\ University, SE-90187 Ume\aa, Sweden}
\affiliation{Institute of Applied Physics, Russian Academy of Sciences, Nizhny Novgorod 603950, Russia}
\author{A.~Bashinov}
\affiliation{Institute of Applied Physics, Russian Academy of Sciences, Nizhny Novgorod 603950, Russia}
\author{I.~Gonoskov}
\affiliation{Department of Physics, Ume\aa\ University, SE-90187 Ume\aa, Sweden}
\author{C.~Harvey}
\affiliation{Centre for Plasma Physics, Queen's University Belfast, BT7 1NN, UK}
\author{A.~Ilderton}
\affiliation{Department of Applied Physics, Chalmers University of Technology, SE-41296 Gothenberg, Sweden}
\author{A.~Kim}
\affiliation{Institute of Applied Physics, Russian Academy of Sciences, Nizhny Novgorod 603950, Russia}
\author{M.~Marklund}
\affiliation{Department of Physics, Ume\aa\ University, SE-90187 Ume\aa, Sweden}
\affiliation{Department of Applied Physics, Chalmers University of Technology, SE-41296 Gothenberg, Sweden}
\author{G.~Mourou}
\affiliation{Institut de la Lumi$\grave{e}$re Extr$\hat{e}$me, ENSTA, Palaiseau, France}
\affiliation{University of Nizhny Novgorod, Nizhny Novgorod 603950, Russia}
\author{A.~Sergeev}
\affiliation{Institute of Applied Physics, Russian Academy of Sciences, Nizhny Novgorod 603950, Russia}

\begin{abstract}
We demonstrate that charged particles in a sufficiently intense standing wave are compressed toward, and oscillate synchronously at, the maxima of the electric field. This unusual trapping behaviour, which we call 'anomalous radiative trapping' (ART), opens up new possibilities for the generation of radiation and particle beams, both of which are high-energy, directed and collimated. ART also provides a mechanism for particle control in high-intensity QED experiments.
\end{abstract}

\maketitle
\paragraph{Introduction:--} Progress in laser technology has opened up possibilities for creating ultra-intense light sources~\cite{Vulcan,ELI,XCELS} with the aim of studying phenomena at the interface of high-field and high-energy physics~\cite{one}. Among these, radiation dominated particle dynamics and quantum electrodynamics effects are of current topical interest and are guiding the direction of upcoming laser programs~\cite{Heinzl:2006xc,Zavattini:2012zs,HIBEF}. A route to achieving the high field strengths needed for studying e.g.\ QED effects is provided by so-called `dipole' pulses~\cite{Ivan}. 

\begin{figure}[t!]
\includegraphics[width=\columnwidth]{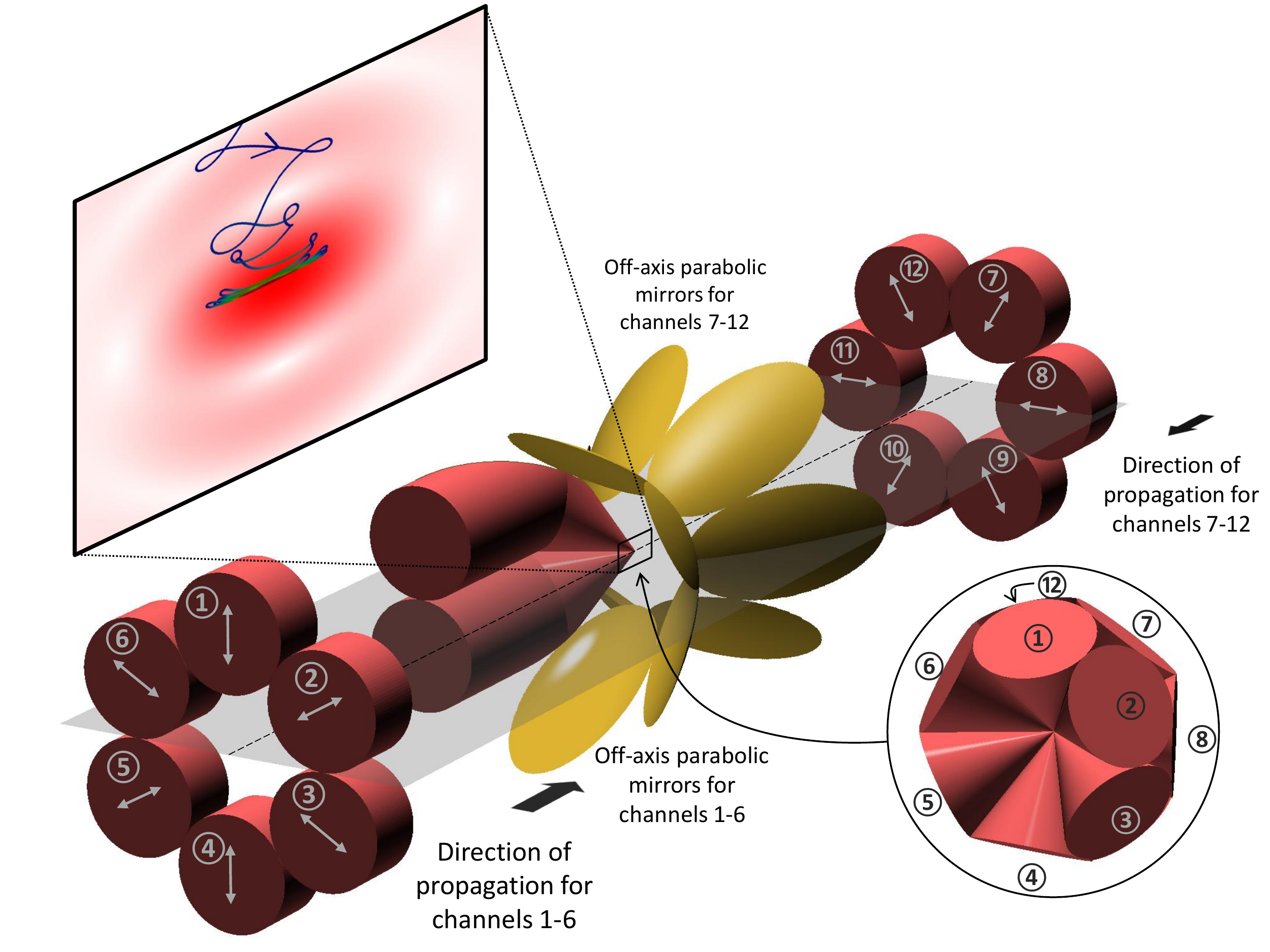}
\caption{Focussing concept for dipole wave production, and the ART effect. Two sets of six counter-propagating beams (polarisation shown with double ended arrows) are reflected by two sets of parabolic mirrors (yellow) aligned such that their surfaces lie along a paraboloid. Both paraboloids have the same symmetry axis (dashed line, $z$-axis) and the same focus point. The resulting focal beam structure is shown in the insert. Top-left: the focal electric field (schematically) and a typical electron trajectory in the ART regime. The particle becomes trapped around the peak of the electric field. \label{12CG}}
\end{figure}

Given fixed input power, the electric field strength in a laser focus can be maximised by using a dipole pulse, which saturates the upper bound on focussing efficiency~\cite{Bassett}. The dipole pulse describes a converging wave of light, which can be pictured as the reverse process of emission from a dipole antenna. Using several channels (e.g.\ as implemented at NIF~\cite{NIF}) to mimic a dipole pulse is the optimal design target for future facilities and offers the potential for going beyond current field strength and intensity records~\cite{Yanovsky:2008}. Fig.~\ref{12CG} shows a focussing concept based on 12 colliding pulses, and which provides $90\%$ of the theoretical maximum electric field strength. (See the appendix for more details.)

In this paper we investigate physics in intense standing waves, such as those provided by the dipole setup. We report the existence of a new regime of particle dynamics in ultra-intense light. We show that charged particles in a sufficiently intense standing wave are compressed toward, and oscillate synchronously at, the maxima of the electric field, rather than the minima. This unusual behaviour, which we call `anomalous radiative trapping' (ART), is due to radiation friction.  We demonstrate in a specific geometry that ART can be used for particle control~\cite{Harvey:2011mp,1} for studying fundamental physics~\cite{Jaeckel:2010ni,DiPiazza:2011tq}, and for the generation of multi-GeV, directed, gamma rays~\cite{two} and collimated, energetic particle beams~\cite{three}.

\paragraph{Particle motion in a dipole wave :--} To investigate particle motion in intense fields, we simulate the relativistic dynamics of (initially uniformly distributed) particles in a converging dipole wave~\cite{Ivan,our-p}. This wave is generated by laser pulses with a Gaussian profile of 30~fs duration (FWHM for intensity), wavelength $\lambda = 810$~nm, and peak total power of 200~PW (averaged over the central period), as is expected to be available at future international projects~\cite{ELI,XCELS}.

Particle motion is due to both the Lorentz force and the particle's own recoil when it radiates, an effect which rises with intensity. Our code contains a classical particle pusher, propagating electrons according to the Lorentz equation. Emission and recoil are implemented at each time step using the quantum theory via statistical routines, using inverse sampling. See~\cite{Elkina:2010up} for a description of the event generator, and \cite{RitusReview} for the probability of emission in ultra-intense fields. This approach is common in particle-in-cell (PIC) codes used for modelling QED processes such as cascades~\cite{Bell:2008zzb,Elkina:2010up}. We neglect secondary QED processes here, as well as the Coulomb force, but the relevance of the latter we will address below.

\begin{figure}[t!]
\includegraphics[width=\columnwidth]{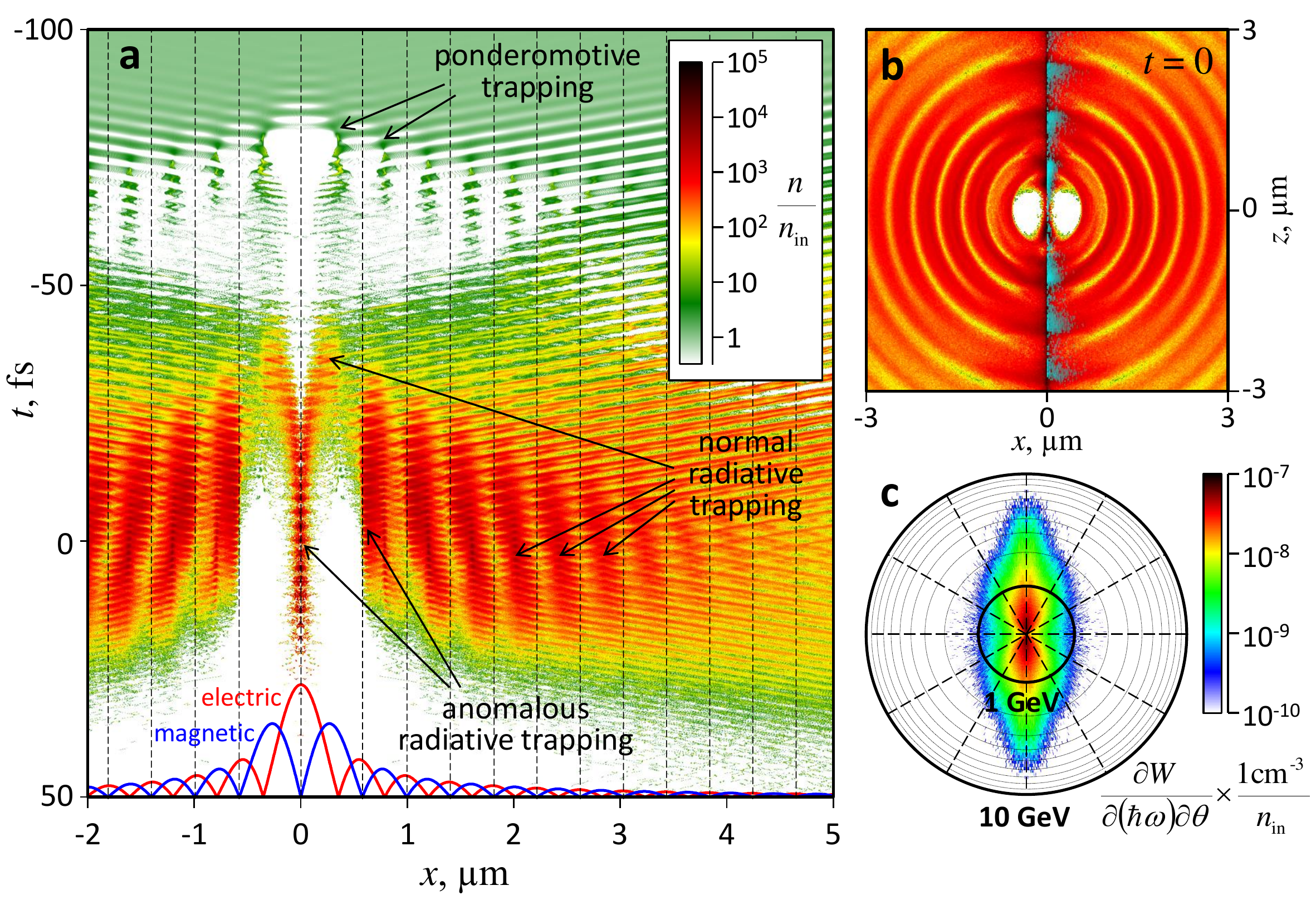}
\caption{Simulation results for electron motion in the dipole wave. \textbf{a}, time evolution of the electron density (divided by the initial density) on the $x$-axis, $z = 0$. Peak electric field locations are shown with dashed lines. \textbf{b}, density distribution at the instance of peak field strength; photons with energy exceeding 3 GeV are shown in cyan on the right hand side. \textbf{c}, photon emission distribution as a function of angle and energy (radial coordinate, log scale). \label{dipole_art}}
\end{figure}

In fig.~\ref{dipole_art}~(a) we plot the time evolution (top to bottom) of the electron density in the focus, $z=0$, as a function of transverse position $x$. (There is rotational symmetry about the $z$-axis, see fig.~\ref{12CG}.) One sees immediately the accumulation and trapping of electrons in different regimes. When the front edge of the pulse reaches the centre it first forms a standing-wave with moderate amplitude. As a result electrons become trapped in the minima of the ponderomotive potential  (describing the average effect of the Lorentz force), coinciding with the positions of the electric field minima. Due to relativistic effects, electrons are released from this {\it ponderomotive trapping} as the field amplitude rises~\cite{UPHILL,LS}. As the role of radiation losses increases, it is known that the particles subsequently become trapped once more, and in the same positions~\cite{Kirk:2009vk}. We call this effect normal radiative trapping (NRT). But remarkably, beginning around $t = 0$, the moment of maximum electric field, the electrons become focussed toward and trapped around the positions of the electric field {\it maxima}, i.e.\ at the maxima of the ponderomotive potential; fig.~\ref{dipole_art}~(b) shows the density distribution at the instance of peak field strength ($t = 0$). We call this counter-intuitive behaviour \textit{anomalous radiative trapping} (ART). We first outline some potential applications of ART, before considering its physical origins.

Our simulations indicate that, even for uniform initial density, the ponderomotive force causes the front edge of the converging dipole wave to push particles toward the centre of the pulse, where large numbers of them are trapped by ART. We refer to electrons trapped in the vicinity of a particular magnetic field node as being in a particular `trapping state'. The number of particles in each trapping state can in fact be controlled by the shape of the pulse's front edge and by the initial particle distribution. The spatial structure of the wave is such that the only channel for electrons to leave the trapping states is along, and in the vicinity of, the $z$-axis. As well as this well collimated source of highly energetic electrons, ART also provides a novel source of well collimated hard photons. In fig.~\ref{dipole_art}~(c) we plot the emitted photon distribution, with energies extending up to 6~GeV (the maximum electron energy). One can distinguish the most energetic peak with energies above 3~GeV and an angular spread of about 10$^{\circ}$. These photons, also shown in fig.~\ref{dipole_art}~(b), are emitted by electrons in the central trapping state.

The maximum number of electrons which could populate the central state is limited by their mutual Coulomb interaction. As a rough estimate for this number, we equate the amplitude of the dipole wave with the total Coulombic field strength of $N$ electrons at the distance of the typical spatial spread of particles in this state (0.1~$\mu$m).  This gives $N_{max} \sim 10^{11}$. Using data from fig.~\ref{dipole_art}~(b, c) we can then estimate the maximum number of photons in the 1~GeV range emitted by these electrons as $N_{h} \sim 10^{12}$,  corresponding to a total energy (100~J) of order 1\% of the initial laser energy.

\begin{figure*}[t!]
\centering\includegraphics[width=\textwidth]{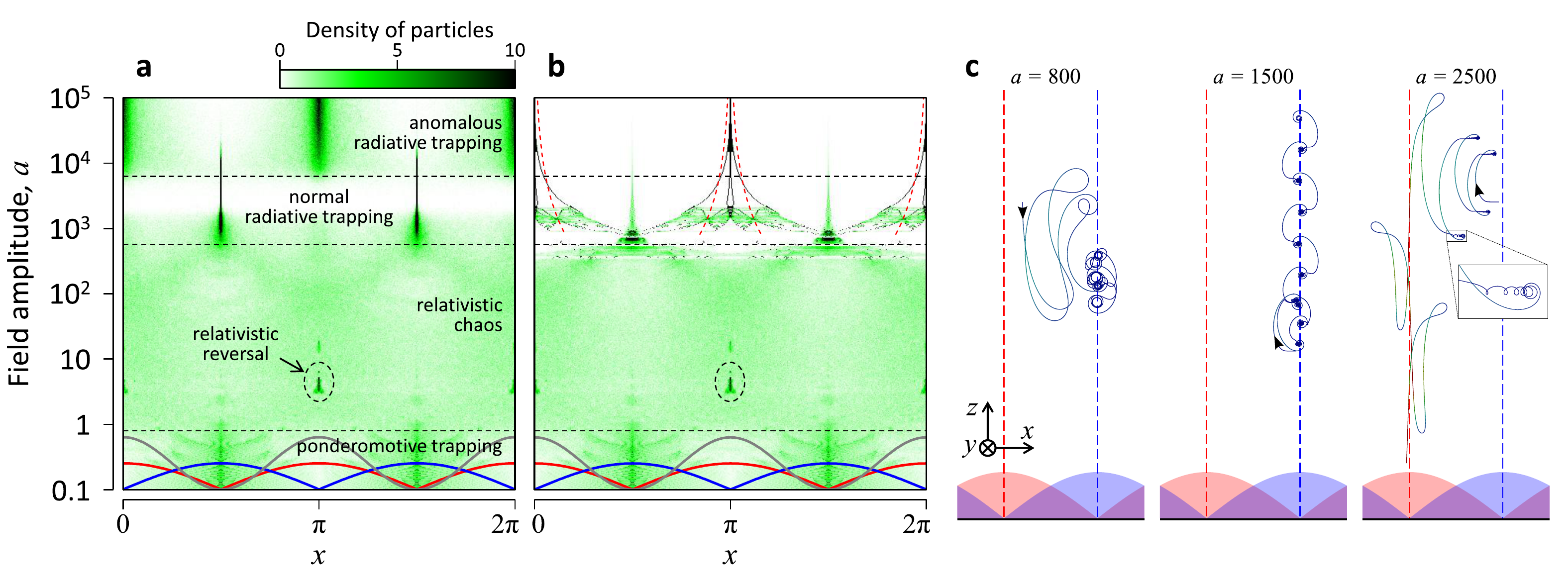}
\caption{
\textbf{a,} The long-term density distribution of electrons in a standing wave as a function of wave amplitude $a$. The spatial distribution of the electric and magnetic fields, and the ponderomotive potential, are sketched with red, blue and grey lines, respectively. Radiation reaction is included via quantum emission. \textbf{b,} The same density distribution calculated using classical radiation reaction. 
\textbf{c,}  Typical particle trajectories. The first and third are in the NRT and ART regimes respectively, while the second shows the transition between them. Dotted red (blue) lines show the peak locations of the electric (magnetic) field.
\label{theory}}
\end{figure*}

\paragraph{Anomalous radiative trapping:--} To understand the basic physics behind ART, we turn to the simpler model of particles in a plane standing wave. From here on position $x$, time $t$ and field strength $a$ are given in units of $\lambda/2\pi$, $\lambda/2\pi c$ and $2\pi mc^2/e\lambda$ respectively.  To assess the relevance of quantum effects, we performed two simulations, calculating the long-term spatial distribution of initially uniformly distributed electrons, for different wave amplitudes.  The first simulation used the same approach as above, that is emission was treated quantum mechanically. The results are shown in fig.~\ref{theory}~(a). The second simulation was entirely classical, with the Landau Lifshitz equation used to describe a radiating particle. The results are shown in fig.~\ref{theory}~(b). The standing wave simulations were started with a uniform distribution of particles, and the long-term distribution was extracted, at the instant of vanishing electric field, after 100 oscillations of the standing wave, when it was observed that the particle distribution had stabilised.

Fig.~\ref{theory}~(a) (the fully quantum simulation) shows ponderomotive trapping, relativistic chaos~\cite{UPHILL,LS}, relativistic reversal~\cite{REVERSE,Dodin}, and the two radiative trapping effects, NRT~\cite{Kirk:2009vk} and ART. The uniform electric (magnetic) field of the standing wave is orientated along the $z$ ($y$)-axis and is shown schematically in red (blue). Comparing with fig.~\ref{theory}~(b) we see that quantised emission causes a broadening of the particle distributions, but that NRT and ART are present both with and without quantum effects. We therefore proceed to use classical radiation reaction to explain NRT and ART.

We first note the dominant role of the magnetic field in causing radiation losses. In the ultra-relativistic limit, these are determined predominantly by the particle's acceleration transverse to it's velocity. The magnetic component of the Lorentz force is always transverse to velocity, whereas the electric component accelerates parallel to the electric field. Thus, its contribution to transverse acceleration depends on the relative orientation of the field and particle velocity. On average, the rate of radiative loss is therefore higher in the vicinity of magnetic field maxima. In the NRT regime this causes the particles to lose their energy and rotate close to the magnetic field maxima, see the first trajectory in fig.~\ref{theory}~(c).

In the NRT regime radiation losses play a small role, and lead to the particle spiralling around the magnetic field maxima in a rather irregular form. As the wave amplitude rises, radiation losses increase and the particles lose essentially all of their energy whenever the magnetic field peaks. Thus, during every temporal maximum of the electric field, the particles are accelerated almost parallel to the electric field, see the second trajectory in fig.~\ref{theory}~(c). This `radiation dominated motion'~\cite{PPR,Bulanov:2010gb} is regular (and iterative) and with rising field amplitude causes net migration toward the spatial maxima of the electric field, see the final trajectory of fig.~\ref{theory}~(c), by the following mechanism.

During periods of acceleration by the electric field, the magnetic field turns particles toward the $\E \times \BB$ direction, while the simultaneous presence of the electric field leads to a drift in the $\E \times \BB$ direction. In a standing wave, $\E \times \BB$ is orientated toward the electric (magnetic) field maximum while the electric field is rising (falling). Thus the particle is shifted toward the electric (magnetic) field maximum when the electric field is rising (falling); call these times stages one and two. 

\begin{figure}[t!]
\includegraphics[width=\columnwidth]{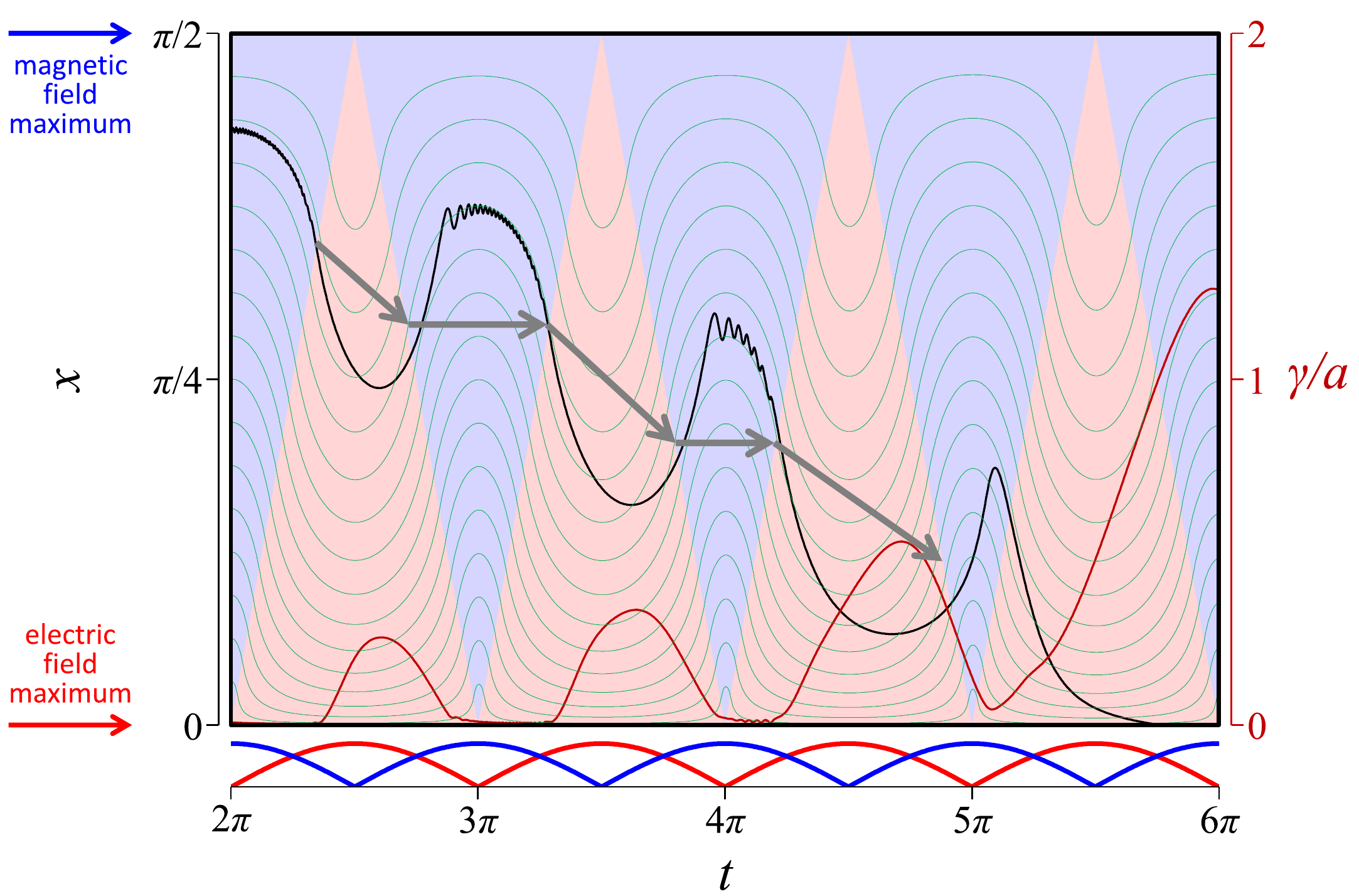}
\caption{Particle trajectory (black) and gamma factor (red) in the ART regime. Red (blue) regions correspond to electric (magnetic) field dominance, $E>B$ ($B>E$). Green lines describe the low energy limit, in which particles mainly gyrate with drift velocity $c\E\times \BB/B^2$ in the blue regions and move essentially linearly in the red regions, with velocity going to $c\E\times\BB/E^2$. In the ART regime, there is no net change in position when the magnetic field dominates. When the electric field dominates there is a clear migration toward the electric field maximum $x=0$.\label{ART_theory}}
\end{figure}

If the particle's dynamics were symmetric with respect to the rising/falling of the electric field, both shifts would be equal.  But because the particle gains energy due to electric field and, as explained above, loses it mostly due to the magnetic field, the gamma factor is larger in stage two than in stage one. A higher gamma factor during stage two means that the particle resists the magnetic field, so the shift away from the electric field maximum is smaller in stage two than the shift toward the maximum in stage one. See fig.~\ref{ART_theory}. Thus, particles migrate toward the electric field maximum in the ART regime within a few oscillations of the standing wave.

When particles reach the vicinity of the electric field maximum they all sit on stable attractors. The typical spatial spread of particles can be estimated as the distance $x_r$ at which the magnetic field can drag particles away from the maxima. This can be roughly estimated as the diameter of rotation in the 
magnetic field. The typical magnetic field strength during rotation
can be estimated as $a\sin x_r \simeq a x_r$. At this distance the magnetic field dominates over the electric for a time interval of roughly $2 x_r$, which we use as an estimate for the time of rotation. Assuming a half turn during this
time interval gives us the relation $\pi \gamma_r /2 \approx x_r^2 a$, where $\gamma_r$ is the typical gamma factor during rotation. We then
assume that up to phase of rotation electric field is roughly balanced by the
radiation reaction force, consequently
\be
	x_r a \approx \frac{4\pi}{3} \frac{r_e}{\lambda}\gamma_r^2 x_r^2 a^2 \;,
\ee
where we suppose both electric and magnetic field are equal to $x_r
a$. In such a way we obtain
\be\begin{split}
	x_r 
	&\approx 0.9
\bigg(\frac{\lambda}{a^3 r_e}\bigg)^{1/5} \;,
\end{split}
\ee
as a rough estimate for the particle spread. This estimate, shown at the top of fig.~\ref{theory}~(b) with dashed red lines, clearly fits the numerical results well, and explains why particles are concentrated toward the electric field maxima with rising amplitude.

The classical equations of motion have, in the ultra-relativistic regime, a similarity parameter $\delta = \left(r_e/\lambda\right)a^3$~\cite{Bulanov:2010gb} defining the transition between relativistic stochastic motion~\cite{LS} and the regimes of NRT and ART. Based on the data of fig.~\ref{theory}~(b) we can identify the threshold values of $\delta$ for both regimes, $\delta_{th}^{NRT} \approx 0.5$, $\delta_{th}^{ART} \approx 600$, corresponding to threshold intensities in terms of $I=(c/8\pi) E_{\text{max}}^2$:
\begin{equation}\label{thresholds}
\begin{array}{l}
	\displaystyle{I_{th}^{NRT} \approx 5 \times 10^{23} \frac{W}{\text{cm}^2} \times \left(\frac{0.81 \mu \text{m}}{\lambda}\right)^{\frac{4}{3}}}, \\
	\displaystyle{I_{th}^{ART} \approx 6 \times 10^{25} \frac{W}{\text{cm}^2} \times \left(\frac{0.81 \mu \text{m}}{\lambda}\right)^{\frac{4}{3}}.} \\
	\end{array}
\end{equation}
Experimental demonstration of NRT is possible using a configuration of two counterpropagating pulses~\cite{Kirk:2009vk} or with optimal focussing; this would require a total power of 1-2~PW, which is within the reach of several current and proposed facilities~\cite{Vulcan}.  ART could be demonstrated at proposed international high intensity facilities such as ELI and XCELS, for which the dipole setup provides $I_{\text{max}} \approx 2 \times 10^{26}$ W/cm$^2$.

\paragraph{Conclusions:--} In summary, we have shown that in focussed fields (standing waves) of high intensity, radiation damping causes particles to become trapped in, rather than expelled from, positions of electric field maxima.  This opens up new possibilities for hard photon generation, charged particle acceleration and for studying QED. \\

\acknowledgments
The simulations were performed on resources provided by the Swedish National Infrastructure for Computing (SNIC)
at HPC2N. The authors are supported by the Ministry of Education and Science of the Russian Federation, Agreement No.\ 11.G34.31.0011 (A.G., G.M., A.S.), the Russian Foundation for Basic Research grant No. 12-02-12086 (A.G.), the Swedish Research Council, contracts 2011-4221 (A.I.), 2010-3727 and 2012-5644 (M.M.), the European Research Council contract 204059-QPQV (A.I., M.M.) and EPSRC grant EP/I029206/1--YOTTA (C.H.). 

\appendix

\section*{Appendix: focussing efficiency}\label{S:Focus}
We define the effective intensity $I_\text{max}$ and focussing parameter $\phi$ by $I_{\text{max}}=cE_{\text{max}}^{2}/{8\pi} = \phi^{2} P/({8\pi}\lambda^{2})$. There is a fundamental upper-limit to $\phi$, which for quasi-monochromatic radiation is~\cite{Bassett} $\phi\leq \phi_{d} = {8 \pi}/\sqrt{3} \approx 14.5$. Dipole pulses saturate this upper bound on focussing efficiency~\cite{Ivan}

The proposal in fig.~\ref{12CG} is the result of an optimisation over different numbers of beams, in different configurations, in the mimicking of the dipole field configuration, see fig.~\ref{configs} below. This study was performed with proposed facilities such as XCELS in mind~\cite{XCELS}, which is expected to provide an intensity of around $I_{\text{max}}=cE_{\text{max}}^{2}/{8\pi} = \phi^{2} P/({8\pi}\lambda^{2}) \approx 2 \times 10^{26}$ W/cm$^2$, assuming 200 PW total power and $\lambda = 810$~nm. We assumed realistic values of f-number not less than unity for the mirrors, and a reasonable number of beams in terms of synchronization~\cite{mourou.nature.photonics}.

Fig.~\ref{configs} shows that using 12 separate laser channels (multi-channel systems are already established at projects such as NIF~\cite{NIF}), ideally aligned~\cite{Ivan} and synchronised~\cite{mourou.nature.photonics}, we can in principle  get to within 90\% of the theoretical maximum field strength for given input power; the focal standing wave in the 12-beam configuration is close to that of the dipole wave, and the focussing parameter is $\phi_{\text{12ch}}\approx 13.0 \approx 90\%\, \phi_{d}$.

\begin{figure}[h!!]
\includegraphics[width=\columnwidth]{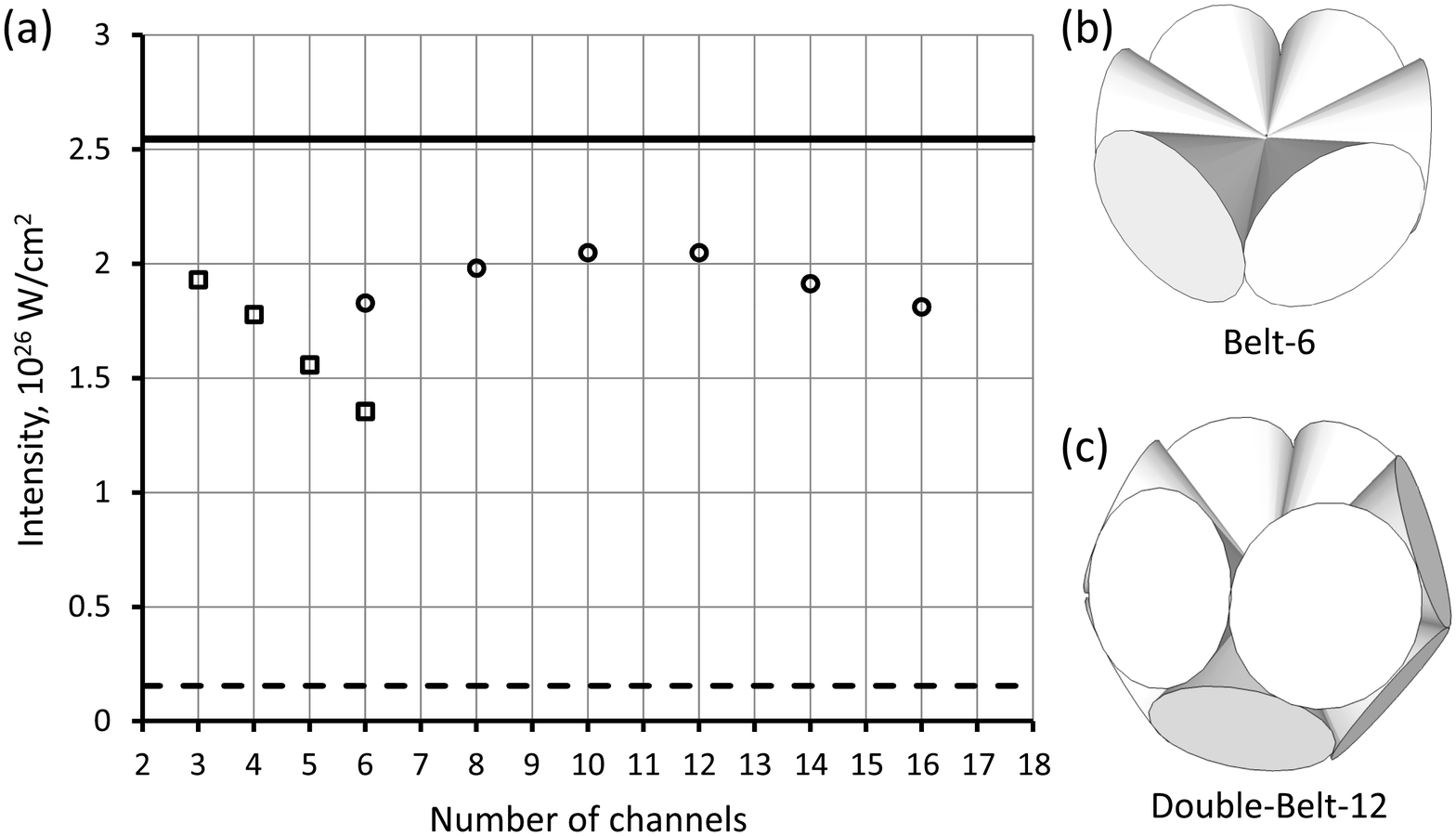}
\caption{\textbf{Dipole wave reconstruction using a finite number of beams.} \textbf{(a)} compares the effective focal intensity in the belt (squares) and double-belt (circles) geometries, for different numbers of channels. The solid line shows the focal intensity for an exact dipole wave, whereas the dashed line shows the focal intensity for single beam focussing with f-number = 1.2 optics. (\textbf{b}) and (\textbf{c}) show examples of the focussed lasers in single and double-belt geometries. \label{configs}}
\end{figure}




\end{document}